\documentstyle[aps,twocolumn,epsfig]{revtex}
\begin{document}
\newcommand{\ve}[1]{\mbox{\boldmath $#1$}}
\twocolumn[\hsize\textwidth\columnwidth\hsize
\newcommand \la{\raisebox{-.5ex}{$\stackrel{<}{\sim}$}}
\newcommand \ga{\raisebox{-.5ex}{$\stackrel{>}{\sim}$}}
\csname@twocolumnfalse%
\endcsname

\draft
\title{Rapidly rotating Bose-Einstein condensates in anharmonic potentials}
\author{G. M. Kavoulakis$^1$ and Gordon Baym$^2$}
\date{\today}
\address{$^1$Mathematical Physics, Lund Institute of Technology, P.O. Box 118,
        S-22100 Lund, Sweden \\
$^2$Department of Physics, University of Illinois at
Urbana-Champaign, 1110 West Green Street, Urbana, Illinois 61801}

\maketitle

\begin{abstract}

    Rapidly rotating Bose-Einstein condensates confined in anharmonic traps
can exhibit a rich variety of vortex phases, including a vortex lattice, a
vortex lattice with a hole, and a giant vortex.  Using an augmented
Thomas-Fermi variational approach to determine the ground state of the
condensate in the rotating frame -- valid for sufficiently strongly
interacting condensates -- we determine the transitions between these three
phases for a quadratic-plus-quartic confining potential.  Combining the
present results with previous numerical simulations of small rotating
condensates in such anharmonic potentials, we delineate the general structure
of the zero temperature phase diagram.

\end{abstract}


\vskip0.0pc]

\section{Introduction}

    Gaseous Bose-Einstein condensates of alkali-metal atoms are fertile
systems in which to study the behavior of superfluids under rotation.  For
rotation frequencies $\Omega$ above $\Omega_{c1}$, the critical frequency for
creating a single vortex, the properties of these systems resemble those of
liquid helium under typical experimental conditions.  States with a small
number of vortices have been observed in the experiments of
Refs.\,\cite{JILA,Madison}.  As $\Omega$ increases, more and more vortices
form, and indeed it has become possible to create lattices
\cite{VortexLatticeBEC,HaljanCornell} of more than 100 vortices.  The physics
of rapidly rotating gases is different in harmonic and anharmonic traps.  In a
harmonic trap, the rotation rate is limited by the radial trap frequency
$\omega$, and sets the scale of $\Omega_{c1}$.  As the angular frequency of
rotation of the gas approaches the transverse trapping frequency, the
centrifugal force becomes equal to the restoring force exerted by the trap and
the atoms are no longer contained.  On the other hand, the regime $\Omega >
\omega$ becomes accessible in anharmonic trapping potentials that rise more
steeply than quadratically.

\begin{figure}[ht]
\begin{center}
\epsfig{file=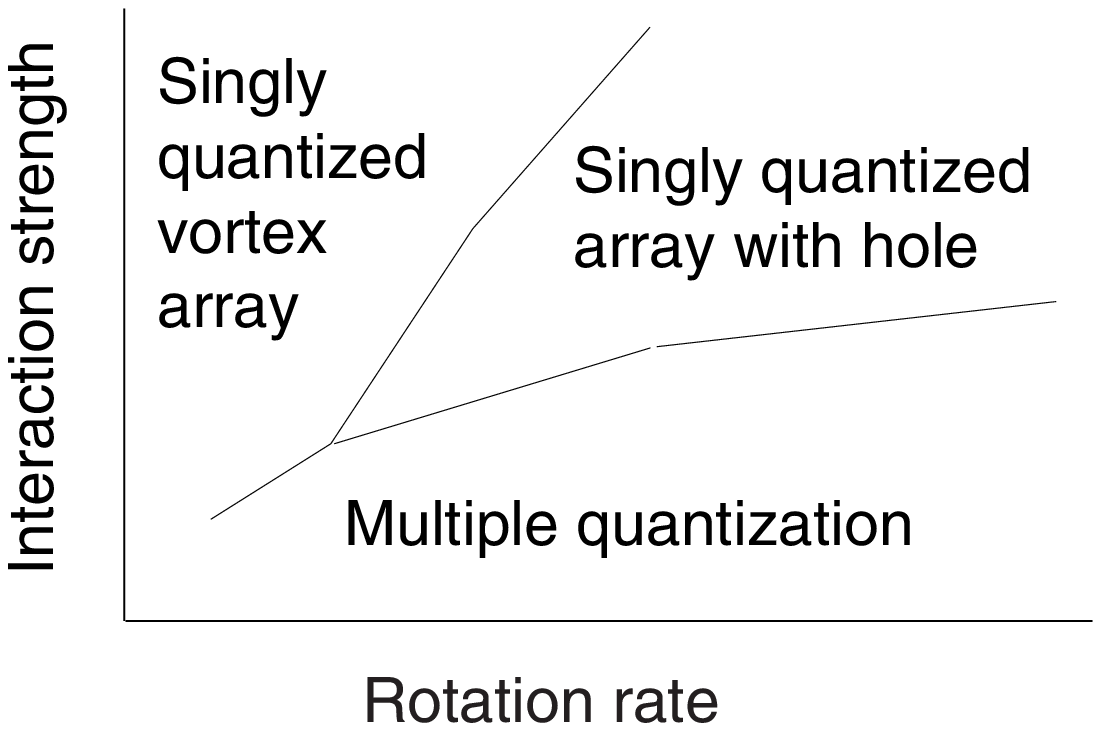,height=4cm}
\end{center}
\vspace{-0.2in}
\begin{caption}
{Schematic phase diagram of the ground state of a rapidly rotating Bose
condensed gas at zero temperature in the $\Omega$--$\sigma a_s$ plane,
showing the regions of multiply quantized vortices, of singly quantized
vortices forming an array which at large number becomes a triangular lattice,
and of an array of singly quantized vortices with a hole in the center.  The
figure does not sketch out the details of the phase diagram at small $\Omega$
below the critical rotation, $\Omega_{c1}$, for onset of vorticity.}
\end{caption}
\label{FIG1}
\end{figure}

    Condensates rotating in anharmonic trapping potentials are expectd to have
a rich structure of vortex states, as described in
Refs.~\cite{Lundh,Fetter,tku,FG}, reviewed below.  These states include an
array of singly quantized vortices (which becomes a triangular lattice when
the number of vortices is large); a vortex array with a finite radius ``hole"
in the center -- opened by the centrifugal force -- in
which the condensate density is zero but the vorticity non-zero; and giant
(multiply quantized) vortices.  Our aim in this paper is first to study, via a
variational approach, the states in an anharmonic quadratic-plus-quartic
radial potential \cite{Kuga},
\begin{equation}
  V(\rho) = \frac{m\omega^2}{2}\rho^2 \left[1 + \lambda
  \left(\frac {\rho}{d}\right)^2\right],
\label{anh}
\end{equation}

    where $\rho$ is the (cylindrical) radial coordinate, $m$ is the atomic
mass, $d=(1/m \omega)^{1/2}$ is the oscillator length ($\hbar = 1$), and
$\lambda$ is a dimensionless measure of the anharmonicity.  We consider a
container infinitely long and cylindrically symmetric and rotating about the
$z$ axis, with a density per unit length $\sigma = N/Z$ in the axial ($z$)
direction; the particles are assumed to interact via s-wave scattering with
scattering length $a_s>0$.  The variational approach is adopted from
Ref.~\cite{FG}, with the density profile of the gas given by an augmented
Thomas-Fermi approximation which explicitly takes into account the important
kinetic energy of flow about the individual vortices and the modification of
the interaction energy by the density variations due to the vortices.  We then
combine the present calculation with the results of the earlier studies to
delineate the phase diagram of a rotating Bose condensate in an anharmonic
trap, when the total vorticity is $\gg$ 1 (i.e., well above the critical
rotation frequency $\Omega_{c1}$).  As we shall see, the phase diagram has the
schematic structure in the $\Omega$--$\sigma a_s$ plane shown in Fig.~1.  At
very small interaction strength or density multiply quantized vortices are
always preferable, as found in \cite{Lundh}, while at larger interaction
strength at small $\Omega$ the system forms an array of singly quantized
vortices.  At larger $\Omega$ the array, as shown in \cite{tku}, develops a
hole in the center.  The phase diagram contains a triple point between the
three regions.

    Lundh \cite{Lundh}, by numerical simulation of the Gross-Pitaevskii
equation in two dimensions, determined the ground state vortex structures in a
rapidly rotating condensate in radial trapping potentials $\sim \rho^n$
($n>2$), as well as in the potential (\ref{anh}) for small total vorticity,
$\lesssim 10$.  His calculations show that for small dimensionless interaction
strength, $\sigma a_s$, the system prefers multiply quantized vortices, as
illustrated in Fig.~1, while above a critical ($\lambda$-dependent)
interaction strength, $(\sigma a_s)_c \sim 1$, the ground state consists of an
array of singly quantized vortices.  The fact that a giant vortex state is
favored at small $\sigma a_s$ can be understood from the fact that in the
limit of a non-interacting gas trapped in a potential that rises faster than
quadratically, the dependence of the single particle energies,
$\epsilon(\nu)$, on angular momentum $\hbar\nu$ rises faster than linear
\cite{Lundh,KMP}.  Thus, in this limit, in the ground state of a fixed number
of bosons in the frame rotating at given $\Omega$ all the particles occupy the
single angular momentum state that minimizes $\epsilon(\nu) - \Omega\nu$.

    Condensates in the potential (\ref{anh}) were also studied via numerical
simulation by Kasamatsu, Tsubota, and Ueda \cite{tku}, at fixed parameters
$\lambda = 1/2$ (corresponding to their $k=1$) and $\sigma a_s = 250/8\pi$.
At low $\Omega$ the system contains an array of singly quantized vortices.
[The value of $\sigma a_s$ in this calculation well exceeds the critical
coupling $(\sigma a_s)_c$ found by Lundh, and is above the triple point in
Fig. 1.] However, as they increase the rotation frequency to $\Omega \simeq
2.14\omega$, the vortex array begins to develop, as a consequence of the
centrifugal force, a hole in the center.  At still higher rotational velocity,
$\Omega \sim 3.2 - 3.5$, the system undergoes a transition to a single
multiply quantized giant vortex.

    Similar behavior was found in Ref.~\cite{FG} for a condensate contained in
a cylindrical hard walled bucket, by constructing an augmented Thomas-Fermi
solution of the Gross-Pitaevskii equation.  At low rotation speed the vortices
form a uniform lattice.  However at a critical rotation frequency, $\Omega_h
\sim 1/2m\xi_0 R$ -- where $\xi_0 = (8\pi{\bar n}a_s)^{-1/2}$ is the coherence
or healing length, and $\bar n$ is the average density -- the lattice begins
to develop a hole in the center.  Finally at a second critical frequency,
$\Omega_G \sim 1/2m\xi_0^2$, the system makes a transition to a giant vortex
state.  Fetter \cite{Fetter}, in considering condensates in the potential
(\ref{anh}), finds a critical rotation speed at which the density in the
center of the trap goes to zero; however, the onset of the holethe is obscured
by the particular approximation employed following factorization of the
condensate wave function.

     While detailed comparison of our calculation with the numerical
simulations of Refs.\,\cite{Lundh} and \cite{tku} is hindered by the fact that
the number of vortices in the simulations is not large enough to form a
uniform lattice, as we assume (see, e.g., Fig.\,1e of Ref.\,\cite{tku}, and
Fig.\,6 of Ref.\,\cite{Lundh}), our results are qualitatively consistent with
and provide a simple theoretical understanding of the numerical results.  At
large $\Omega/\omega$ and $\sigma a_s$ where our approach is valid, we see
that with increasing $\Omega$ the system indeed undergoes a transition from a
vortex lattice to a vortex lattice with a hole around the axis of rotation.
Furthermore, as $\Omega$ increases the system eventually undergoes a
transition to a single giant vortex, for $\sigma a_s$ up to at least 350, the
largest value for which we studied this transition.

    We first describe the three states of the rapidly rotating gas -- a vortex
lattice, both with and without a hole along the axis, in Sec.\,IIA, and a
giant vortex, in Sec.\,IIB -- and calculate the augmented Thomas-Fermi
solutions for the lowest energy in the rotating frame at given $\Omega$.  In
Sec.\,III we determine the transitions between different phases for the
particular anharmonicity $\lambda=1/2$, and the less steep $\lambda=1/8$.  In
Sec.\,IV we examine the limits of validity of our approach, and finally
summarize our results in Sec.\,V.

\section{Possible vortex states of a rapidly rotating gas}

\subsection{Vortex lattice}

    We first consider a uniform lattice of singly quantized vortices.  As
shown in Ref.\,\cite{FG}, the energy of the system in the rotating frame,
$E'_{\rm L}$, in the Thomas-Fermi limit is
\begin{eqnarray}
    E_{\rm L}' &=& \int d^3{r}\, n_{\rm L}({\rho})
  \left[ \left( \frac {\Omega_v^2} 2  - \Omega_v \Omega \right) m \rho^2
      + (a+1) \Omega_v - \Omega
      \right. \nonumber\\ & & \left.
       + V(\rho) + \frac 1 2 U_0 n_{\rm L}({\rho}) b \right],
\label{ErotOmegav}
\end{eqnarray}
where $\Omega$ is the rotational frequency of the trap, $\Omega_v$ is the
rotational frequency of the vortex lattice, and $U_0 = 4 \pi \hbar^2 a_s/m$.
We assume an effective core radius $\xi$ (different in general from the
Gross-Pitaevskii healing length $\xi_0$).  The function $a(\zeta)$ describes
the kinetic energy of rotation of the vortices, and $b(\zeta)$ the effects
of the vortex cores on the interaction energy.  For a simple model in which
the order parameter ramps up linearly over a distance $\xi$ from the cores of
the vortices, and then flattens out, $a$ and $b$ are given by \cite{FG},
\begin{eqnarray}
 a(\zeta) = \frac{1-\ln\sqrt{\zeta}}{1- \zeta/2};
  \, \,\,  b(\zeta) = \frac {1-2\zeta/3} {(1- \zeta/2)^2},
\label{abdef}
\end{eqnarray}
where $\zeta = (\xi/\ell)^2$, and $\ell=1\sqrt{m\Omega_v}$ is the radius
of the cylindrical Wigner-Seitz cell corresponding to a given vortex \cite{FG}.

    Minimization of $E'_{\rm L}$ with respect to $n_{\rm L}({\rho})$ yields
the Thomas-Fermi profile,
\begin{eqnarray}
 n_{\rm L}(\rho) &=&
 \frac 1 {U_0 b} \left[\mu - (a+1) \Omega_v + \Omega  \right.
   \nonumber \\
 &-&  \frac12 m \rho^2
 (\Omega_v^2 - 2 \Omega \Omega_v) - V(\rho)
  \left.\right],
\label{n}
\end{eqnarray}
with $\mu$ the chemical potential.  Since the density of the gas at the
center of the cloud, $\rho = 0$, cannot be negative, the system begins to
develop a hole when $n_{\rm L}(\rho = 0)$ falls to zero, or from
Eq.\,(\ref{n}), $\mu - (a+1) \Omega_v + \Omega = 0$.  For smaller $\Omega$,
this quantity is positive and the inner radius of the cloud, $R_1$, vanishes,
while for larger $\Omega$, $R_1 \neq 0$.  We note that when a hole develops in
a uniform vortex lattice, the hole must, for the lattice to remain uniform,
contain the same vorticity, $\pi R_1^2 n_v = mR_1^2\Omega_v$ (where $n_v$ is
the number of vortices per unit area), as would be present were the hole
filled with condensate.  Equation\,(\ref{n}) implies that the inner and outer
radii of the cloud, $R_1$ and $R_2$, respectively, where the density vanishes,
are given by the solutions of
\begin{eqnarray}
      \frac12{\tilde R}_i^2 \left[1 + \frac12\lambda
       {\tilde R}_i^2\right] + (a+1) \tilde\Omega_v - \tilde\Omega
\nonumber \\
   + \frac{1}{2} {\tilde r}_i^2 (\tilde\Omega_v^2 - 2 \tilde\Omega
  \tilde\Omega_v) = \tilde\mu,
\label{ri}
\end{eqnarray}
where here lengths with a tilde are measured in units of $d$, e.g., ${\tilde
R_i}=R_i/d$, $i=1,2$, and energies or frequencies with a tilde are measured in
units of $\omega$.

    Integrating the density over the volume, $\int n_{\rm L} \, d^3r = N$, we
find,
\begin{eqnarray}
  4 \sigma a_s b =
    \tilde\mu ({\tilde R_2}^2 - {\tilde R_1}^2) - \frac14
 ({\tilde R_2}^4 - {\tilde R_1}^4)
 \nonumber \\
  - \frac{\lambda}{6}({\tilde R_2}^6 - {\tilde R_1}^6)
  + ({\tilde R_2}^2 - {\tilde R_1}^2) [\tilde\Omega - (a+1)
        \tilde\Omega_v]
  \nonumber \\
 - \frac14 ({\tilde R_2}^4 - {\tilde R_1}^4) (\tilde\Omega_v^2 - 2
        \tilde\Omega \tilde\Omega_v).
\label{si1}
\end{eqnarray}
Minimization of $E'_{\rm L}$ with respect to $\Omega_v$ yields
\begin{eqnarray}
  \Omega_v = \Omega - N (a+1) / I,
\label{omevome}
\end{eqnarray}
where $I= \int d^3{r}\, m \rho^2 n_{\rm L}(\rho)$ is the moment of inertia.
Equation~(\ref{omevome}), with Eq.\,(\ref{n}), implies that
\begin{eqnarray}
     8\sigma a_s  b (a+1) = (\tilde\Omega - \tilde\Omega_v) \left[ \tilde\mu
({\tilde R_2}^4 - {\tilde R_1}^4)
\right. \nonumber \\ \left.
- \frac{\lambda}{4} ({\tilde R_2}^8 - {\tilde R_1}^8)
   + ({\tilde R_2}^4 - {\tilde R_1}^4) [\tilde\Omega - (a+1) \tilde\Omega_v]
\right. \nonumber \\ \left.
- \frac13 ({\tilde R_2}^6 - {\tilde R_1}^6)
  (\tilde\Omega_v^2 - 2 \tilde\Omega \tilde\Omega_v) \right].
\label{ei}
\end{eqnarray}
For a given value of the dimensionless parameters $\sigma a_s$, $\lambda$,
and $\zeta$, we solve Eqs.  (\ref{ri}), (\ref{si1}), and (\ref{ei}) to find
$R_1$, $R_2$, $\mu$, and $\Omega_v$; then from Eqs.~(\ref{ErotOmegav}) and
(\ref{n}) we construct the energy $E_{\rm L}'(\Omega, \zeta)$ in the
rotating frame, as a function of $\Omega$ and $\zeta$, and finally vary
$\zeta$ to minimize $E_{\rm L}'$.  The total angular momentum of the gas is
given by \cite{baymchandler}
\begin{equation}
  L_z =  \int d^3{r}\, n_{\rm L}({\rho}) (1 + \Omega_v m \rho^2).
\end{equation}
A useful check on the numerical results is the relation $\partial E_{\rm
L}' / \partial \Omega = - L_z$, implied by Eq.\,(\ref{ErotOmegav}).

\subsection{Giant vortex}

    We next consider the giant vortex state, whose vorticity is
localized along the rotation axis, and order parameter is
\begin{eqnarray}
 \Psi_{\rm G}({\bf r}) = \sqrt{n_{\rm G}(\rho)} \, e^{i \nu \phi},
\end{eqnarray}
where $n_{\rm G}({\rho})$ is the density profile of the cloud, $\rho$ and
$\phi$ are cylindrical polar coordinates, and $\nu$ is the winding number.
The angular momentum per particle along the $z$ axis is $L_z/N = \nu$, while
the energy of the system in the rotating frame in the Thomas-Fermi limit is
\begin{equation}
 E_{\rm G}'
  = \int d^3{r} \,  n_{\rm G}({\rho}) \left( \frac {\nu^2} {2 m \rho^2} +
    V(\rho) + \frac {U_0} 2  n_{\rm G}({\rho})  - \nu \Omega \right).
\label{EGiantnu}
\end{equation}
Setting the variational derivative $\delta E_{\rm
G}'/\delta n_{\rm G}({\rho})$ equal to the chemical potential, $\mu$, we find
the density profile of the cloud:
\begin{eqnarray}
  n_{\rm G}(\rho) = \frac 1 {U_0} [\tilde{\mu}' - \frac {\nu^2} {2 m \rho^2}
    - V(\rho)],
\label{pgv}
\end{eqnarray}
where $\mu'=\mu+\nu\,\Omega$.  The inner and outer radii of the cloud, $R_1$
and $R_2$, solve
\begin{equation}
 \frac {\nu^2} {2 {\tilde R_i}^2} + \frac{{\tilde R_i}^2}{2}
   \left(1 + \frac{\lambda}2 {\tilde R_i}^2\right) =
     \tilde{\mu}',\quad i=1,2.
\label{rgvi}
\end{equation}
In addition, the constraint $\int n_{\rm G} \, d^3r = N$ implies that,
\begin{eqnarray}
     4\sigma a_s = \tilde{\mu}' \left({\tilde R_2}^2 - {\tilde R_1}^2\right)
      &-& \nu^2 \ln \frac{\tilde R_2}{\tilde R_1}
        - \frac14 ({\tilde R_2}^4 - {\tilde R_1}^4)
\nonumber \\
 &-& \frac {\lambda} {6} ({\tilde R_2}^6 - {\tilde R_1}^6).
\label{cont1}
\end{eqnarray}
Minimization of $E_{\rm G}'$ with respect to $\nu$ yields
\begin{eqnarray}
  \nu \int \frac {n_{\rm G}} {m \rho^2} \, d^3 r = N \Omega,
\label{nu1}
\end{eqnarray}
and thus
\begin{eqnarray}
   4 \sigma a_s {\tilde \Omega} = \nu \left[ 2 \tilde{\mu}' \ln \frac
{\tilde R_2}{\tilde R_1}
   + \frac {\nu^2}{2} \left(\frac 1 {\tilde R_2^2} - \frac 1
{\tilde R_1^2}\right)
      \right.
\nonumber \\ \left.
  - \frac12({\tilde R_2}^2 - {\tilde R_1}^2)
 - \frac{\lambda}{4}\left({\tilde R_2}^4 - {\tilde R_1}^4\right)\right].
\label{cond4}
\end{eqnarray}
For given values of the dimensionless parameters $\sigma a_s$ and
$\lambda$, we solve Eqs.  (\ref{rgvi}), (\ref{cont1}), and (\ref{cond4}), and
express $R_1$, $R_2$, $\mu'$ and $\Omega$ in terms of $\nu$.  We then
numerically solve for $E_{\rm G}'$ from Eqs.~(\ref{EGiantnu}) and (\ref{pgv})
as a function of $\nu$, and thus implicitly as a function of $\Omega$.  The
relation $\partial E_{\rm G}'/\partial \Omega = - L_z$, implied by
Eq.\,(\ref{EGiantnu}), again provides a check on the numerical results.

\section{Results}

    We have limited the calculations of the vortex phases here to the cases of
intermediate anharmonicity, $\lambda = 1/2$, and weak anharmonicity, $\lambda
= 1/8$, which well illustrate the richness of the phase diagram.  Figure 2
shows the phase diagram for $\lambda=1/2$.  Relatively small $\Omega$ and
relatively large $\sigma a_s$ favor the formation of a vortex lattice.  In the
opposite limit, large $\Omega$ and small $\sigma a_s$ favor the formation of a
giant vortex.  Finally at intermediate $\Omega$ (for $\sigma a_s$ in the range
considered) the system prefers a vortex lattice with a hole.  Quite generally,
the transition from a vortex lattice to a vortex lattice with a hole is
continuous, second order, while the transition from a vortex lattice with or
without a hole to a giant vortex is discontinuous, first order.

\begin{figure}
\begin{center}
\epsfig{file=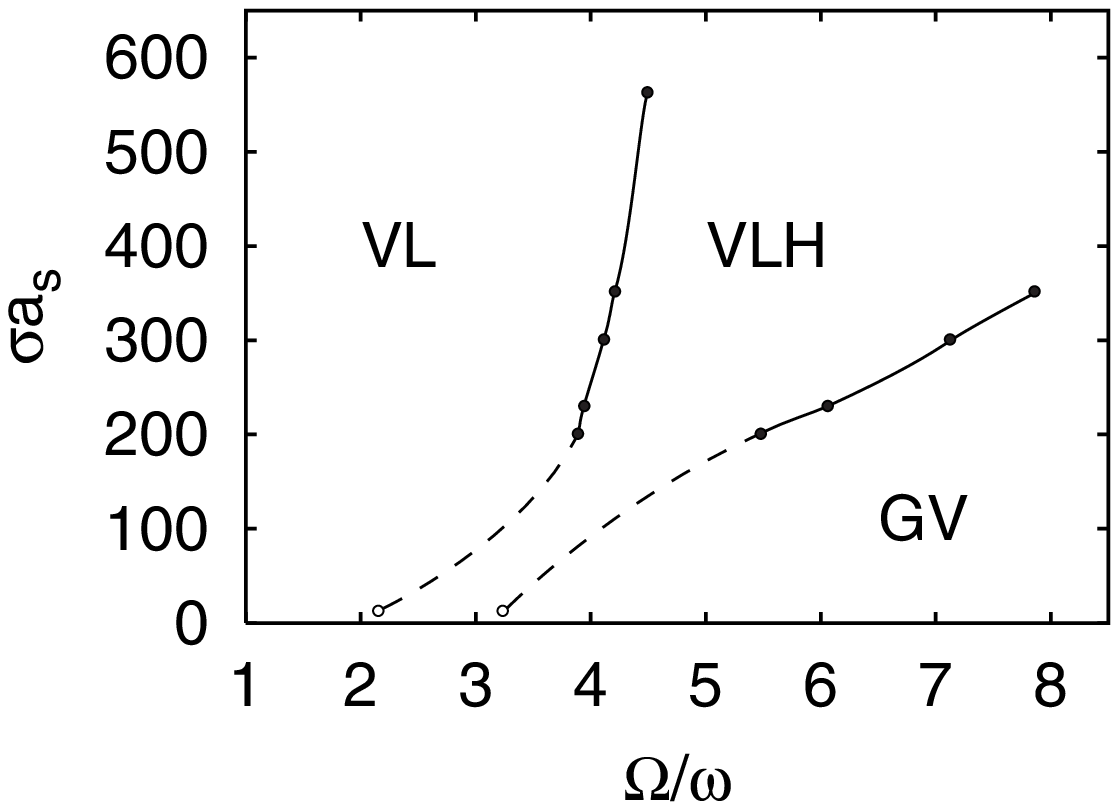,width=8.0cm,height=5.5cm,angle=0}
\end{center}
\begin{caption}
{The phase diagram of the rapidly rotating Bose condensed gas at zero
temperature in the $\Omega$--$\sigma a_s$ plane, for anharmonicity
$\lambda=1/2$.  The labels VL, VLH and GV denote the regions where the favored
state is a vortex lattice, a vortex lattice with a hole, and a giant vortex,
respectively.  The filled circles at $\sigma a_s\ge 200$
indicate the interaction strengths explicitly calculated here, while the two
open circles at $\sigma a_s \sim 10$ show the calculations of Ref.~[7].
The lines are a guide to the eye.}
\end{caption}
\label{FIG2}
\end{figure}

    For $\sigma a_s = 200$, the vortex lattice develops a hole for
$\Omega/\omega \approx 3.90$, and then makes a transition to a giant vortex
for $\Omega/\omega \approx 5.48$.  At this point the giant vortex has a
vorticity $\nu \simeq 168$.  For $\sigma a_s = 550$, corresponding to the
parameters of the multivortex MIT experiment Ref.\,\cite{VortexLatticeBEC}, we
find that the vortex lattice develops a hole for $\Omega/\omega \approx 4.50$,
and that even up to the highest value of $\Omega/\omega \approx 8.4$ that we
examined, this solution has lower energy than the giant vortex.  For such high
values of $\sigma a_s$ and $\Omega$ we observe that the two solutions (vortex
lattice with a hole and giant vortex) look macroscopically very similar.

    For comparison, we have also calculated the phases for interaction
strength $\sigma a_s = 250/8\pi \approx 10$ considered in Ref.\,\cite{tku},
although it is well below the limits of validity of the present approach.
Then the transition to a giant vortex occurs for $\Omega/\omega \approx 1.5$,
compared with the value $\sim$ 3.2 found in Ref.\,\cite{tku}; however $\nu$
(taken to be continuous in the calculation) is between 2 and 3. For such low
values of $\nu$ our approach is clearly not valid, and the discrepancy should
not be taken seriously.

\begin{figure}
\begin{center}
\epsfig{file=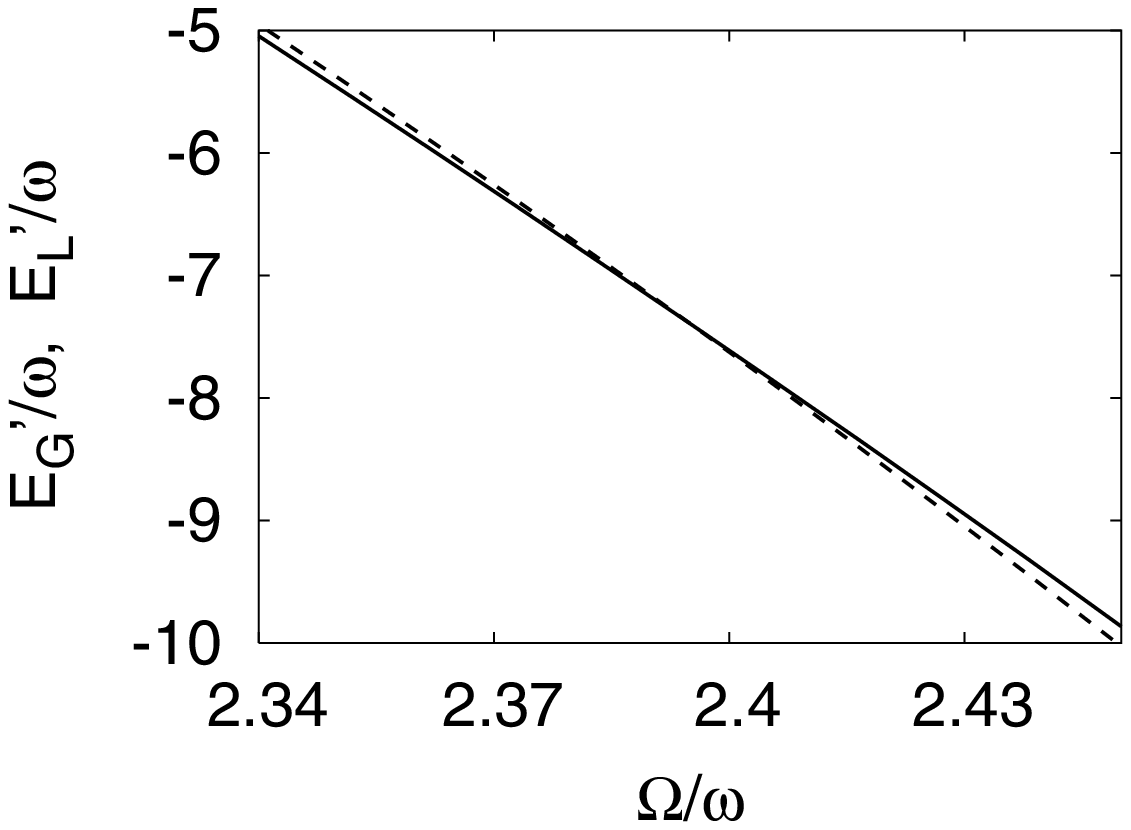,width=8.0cm,height=5.5cm,angle=0}
\begin{caption}
{The energies in the rotating frame of the giant vortex, $E'_{\rm G}$
(dashed curve) and of the vortex lattice, $E'_{\rm L}$ (solid curve) as a
function of $\Omega$, for anharmonicity $\lambda = 1/8$, and
dimensionless interaction strength $\sigma a_s = 90$.}
\end{caption}
\end{center}
\label{FIG3}
\end{figure}

\begin{figure}
\begin{center}
\epsfig{file=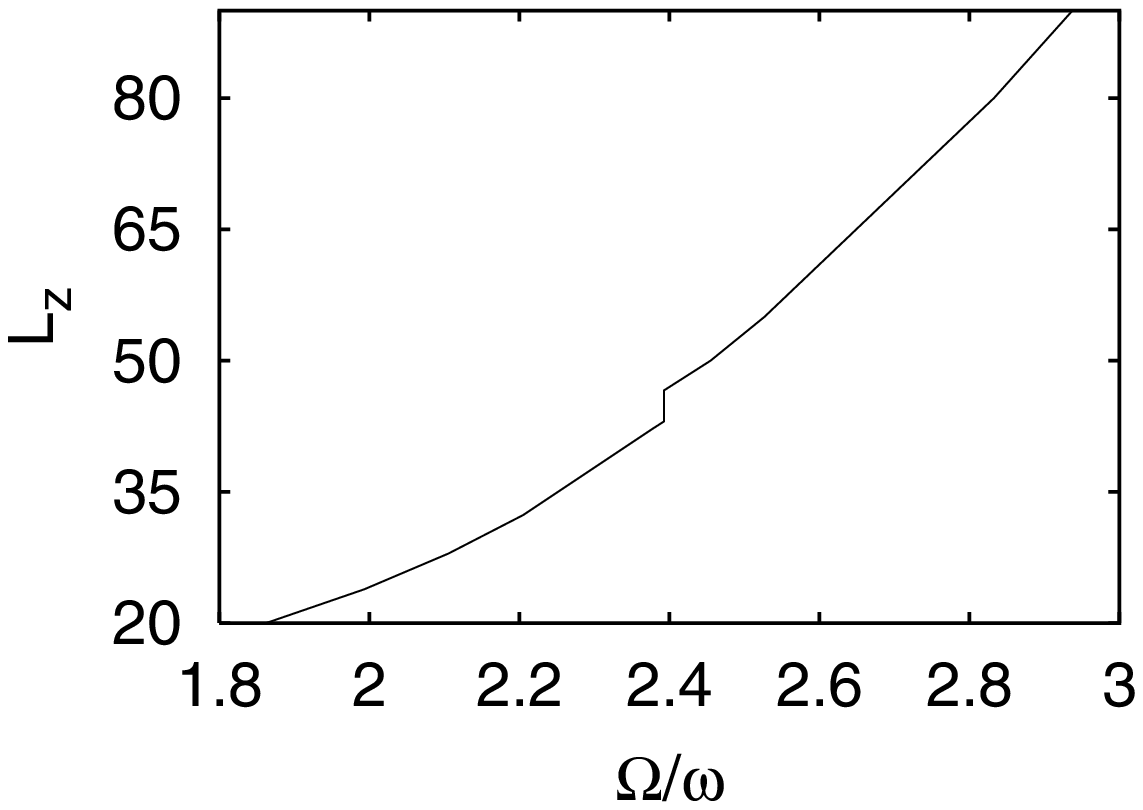,width=8.0cm,height=5.5cm,angle=0}
\begin{caption}
{The angular momentum per particle, $L_z/N$, versus the angular frequency
$\Omega$, for $\lambda = 1/8$ and $\sigma a_s = 90$.}
\end{caption}
\end{center}
\label{FIG4}
\end{figure}

    The particular case of weak anharmonicity case $\lambda = 1/8$, with
$\sigma a_s = 90$ also shows the three phases:  for $\Omega/\omega \approx
2.30$ the vortex lattice develops a hole in the middle, while for
$\Omega/\omega \approx 2.40$ it makes a transition to a giant vortex.  Figure
3 shows the gentle crossing of $E_{\rm G}'(\Omega)$ (solid curve), and $E_{\rm
L}'(\Omega)$ (dashed curve), while Fig.\,4 shows the angular momentum per
particle $L_z/N$ (of the lowest state), the negative of the slope of the lower
of the two curves of Fig.\,3, as a function of $\Omega$.  Since the curves
cross at a finite but small angle, $L_z/N$ is discontinuous at the point of
the transition.  Figure\,5 shows $R_1$ and $R_2$ as a function of $\Omega$ for
the three solutions.

\begin{figure}
\begin{center}
\epsfig{file=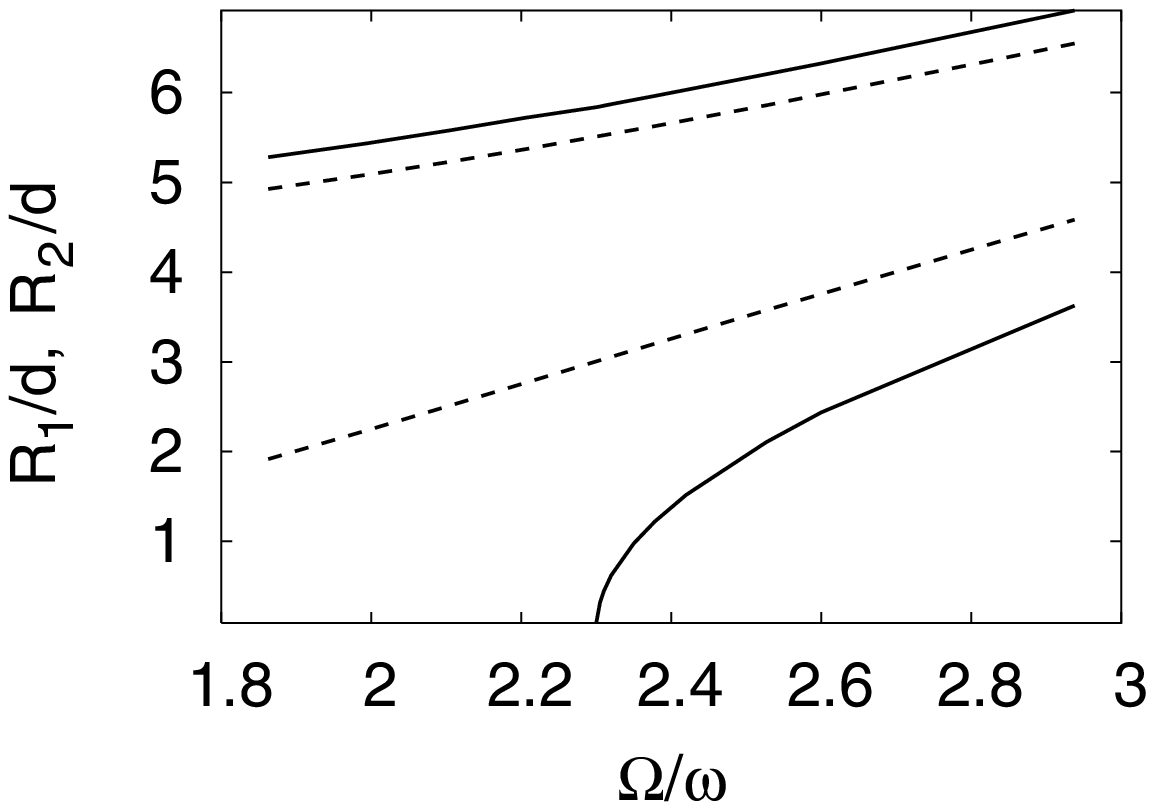,width=8.0cm,height=5.5cm,angle=0}
\begin{caption}
{The inner and outer radii, $R_1$ and $R_2$, of the cloud vs. the angular
frequency $\Omega$, for $\lambda = 1/8$, and interaction
strength $\sigma a_s = 90$.  The solid curves are for the vortex lattice, and
the dashed ones for the giant vortex.  As the lowest curve illustates, the
vortex lattice develops a hole for $\Omega/\omega \approx 2.30$.  The
transition to the giant vortex takes place at $\Omega/\omega \approx 2.40$.}
\end{caption}
\end{center}
\label{FIG5}
\end{figure}

\section{Limits of validity of the present approach}

    The present approach makes two crucial assumptions, first that the profile
of the gas can be described by the Thomas-Fermi approximation, with the vortex
corrections to the kinetic and interaction energies included, and second that
the array of singly quantized vortices is a uniform lattice.  The validity of
the Thomas-Fermi approach requires that the kinetic energy per particle, $\sim
1/2 m (R_2-R_1)^2$, associated with localizing the atoms within the radii of
the profile of the cloud $R_1$ and $R_2$, be much less than $\bar n U_0 =
1/2m\xi_0^2$,
where $\bar n \sim N/\pi(R_2^2 - R_1^2)Z$ is the average density of the cloud,
and $\xi_0$ is the coherence length.  Thus validity of Thomas-Fermi requires
\begin{eqnarray}
 8\sigma a_s \gg \frac {R_1 + R_2} {R_2 - R_1},
\label{in}
\end{eqnarray}
or equivalently, $\xi_0 \ll R_2-R_1$.  For $R_1 = 0$, the condition
(\ref{in}) is simply $8 \sigma a_s \gg 1$.

    In addition to condition (\ref{in}), the basic features of the lowest
state have to be captured by the trial functions being considered.  For small
values of $\sigma a_s$ and $\Omega/\omega$, the numerical solutions
\cite{Lundh,tku} have too small a vorticity to arrange in a triangular
lattice.  Rather, in Ref.  \cite{tku} at $\Omega/\omega = 2.5$ for example,
the vortex hole, containing 4 units of vorticity, is surrounded by a ring of
10 vortices, instead of a developed lattice.  A reasonable estimate for the
lowest value of $N_v$ at which the calculation is valid is $\sim$ 50, which
typically (e.g., at $\lambda = 1/2$) requires $\Omega/\omega \agt 3.5$.

\section{Summary}

    In the present study we have examined, by means of a variational method,
the lowest state of a Bose-Einstein condensate rotating in a
quadratic-plus-quartic confining potential, in the limit of fast rotation.  We
find that as the frequency of rotation, the density of the gas, and the
strength of the quartic term in the trapping potential are varied, the system
undergoes transitions among three phases:  a vortex lattice with and without a
hole, and a giant vortex.  The phase diagram, whose structure we have mapped
for a specific choice of the coefficient of the anharmonic term, is quite
rich.  This structure is not limited to the specific form of the anharmonic
potential considered here.  Our approach is reliable in the limit of large
densities and high angular frequencies.  Many problems remain, including
determining the details of the critical point in Fig. 1, the variation of the
phase diagram as the coefficient of the quartic term in the confining
potential is varied, and finally, the structure of the phase diagram at finite
temperatures.

\vskip1.0pc

\noindent
    Author GMK acknowledges financial support from the Swedish Research
Council (VR), and from the Swedish Foundation for Strategic Research (SSF).
This research was supported in part by NSF Grant PHY00-98353.

\end{document}